\documentclass[onecolumn]{aa} 

\usepackage{graphicx}
\usepackage{color}
\usepackage{txfonts}
\begin{document}

   \title{Optical imaging for the {\it Spitzer} Survey of Stellar Structure in Galaxies\thanks{Tables~\ref{CDStable} and 3 and FITS files of the images  are available at the Centre de Donn\'res astronomiques de Strasbourg (CDS) via anonymous ftp to cdsarc.u-strasbg.fr (130.79.128.5) or via http://cdsarc.u-strasbg.fr/viz-bin/qcat?J/A+A/xxx/yyy. The images are also offered through the NASA/IPAC Extragalactic Database (NED).}}

   \subtitle{Data release and notes on interacting galaxies}

   \author{Johan H. Knapen\inst{1,2}
          \and
          Santiago Erroz-Ferrer\inst{1,2}
          \and
          Javier Roa\inst{1,3}
          \and Judit Bakos\inst{1,2,4}
          \and Mauricio Cisternas\inst{1,2}
          \and Ryan Leaman\inst{1,2}
          \and Nik Szymanek\inst{5}
          }

   \institute{Instituto de Astrof\'\i sica de Canarias, E-38200 La Laguna, Spain; 
              \email{jhk@iac.es}
         \and
             Departamento de Astrof\'\i sica, Universidad de La Laguna, E-38206 La Laguna, Spain
         \and 
             Space Dynamics Group, Universidad Polit\'ecnica de Madrid, E-28040 Madrid, Spain
         \and 
         Konkoly Observatory, Research Centre for Astronomy and Earth Sciences, Hungarian Academy of Sciences, Konkoly Thege Mikl\'os \'ut 15-17, H-1121 Budapest, Hungary
         \and
             186 Thorndon Avenue, West Horndon, Essex, CM13 3TP, UK 
             }

   \date{Received; accepted 3 June 2014}

 
  \abstract
   {The {\it Spitzer} Survey for Stellar Structure in Galaxies (S$^4$G) and its more recently approved extension will lead to a set of 3.6 and 4.5$\mu$m images for 2829 galaxies, which can be used to study many different aspects of the structure and evolution of local galaxies.}
   {We have collected and re-reduced optical images of 1768 of the survey galaxies, aiming to make these available to the community as ready-to-use FITS files to be used in conjunction with the mid-IR images. Our sky-subtraction and mosaicking procedures were optimised for imaging large galaxies. We also produce false-colour images of some of these galaxies to be used for illustrative and public outreach purposes.}
   {We collected and re-processed images in five bands from the Sloan Digital Sky Survey for 1657 galaxies, which are publicly released with the publication of this paper. We observed, in only the $g$-band, an additional 111 S$^4$G galaxies in the northern hemisphere with the 2.5\,m Liverpool Telescope, so that optical imaging is released for 1768 galaxies, or for 62\% of the S$^4$G sample. We visually checked all images. We  noted interactions and close companions in our optical data set and in the S$^4$G sample, confirming them by determining the galaxies' radial velocities and magnitudes in the NASA-IPAC Extragalactic Database.
   }
   {We find that 17\% of the S$^4$G galaxies (21\% of those brighter than 13.5\,mag) have a close companion (within a radius of five times the diameter of the sample galaxy, a recession velocity within $\pm200$\,km\,s$^{-1}$ and not more than 3 mag fainter) and that around 5\% of the bright part of the S$^4$G  sample show significant morphological evidence of an ongoing interaction. This confirms and further supports previous estimates of these fractions. }
   {The over 8000 science images described in this paper, the re-processed Sloan Digital Sky Survey ones, the new Liverpool Telescope images, the set of 29 false-colour pictures, and the catalogue of companion and interacting galaxies, are all publicly released for general use for scientific, illustrative, or public outreach purposes.}


   \maketitle
%

\section{Introduction}

The {\it Spitzer} Survey of Stellar Structure in Galaxies (S$^4$G; Sheth et al. 2010) has used the {\it Spitzer Space Telescope} Infrared Array Camera (IRAC; Fazio et al. 2004) at 3.6 and 4.5\,$\mu$m to observe the stellar mass distribution in a sample of 2352 galaxies\footnote{The size of the original S$^4$G sample as described in Sheth et al. 2010 was 2331 galaxies, but we refer here to 2352 the galaxies actually observed in the survey and released through the NASA/IPAC Infrared Science Archive (IRSA).}. These bands are excellent tracers of the underlying stellar mass distribution, from which the gravitational potential can be derived, especially once they have been corrected for the limited contributions of young stars (e.g., Meidt et al. 2012; Eskew et al. 2012; Meidt et al. 2014, M. Querejeta et al. in preparation). A recently approved extension to the original S$^4$G survey will add 477 galaxies\footnote{The sample of the S$^4$G extension proposal for {\it Spitzer} has 695 galaxies, but 218 of these overlap with the original S$^4$G.} of predominantly early morphological types, for which 3.6 and 4.5\,$\mu$m are being obtained in the current Cycle 10 of {\it Spitzer} observations. This will bring the size of the combined S$^4$G  sample to a total of 2829 nearby galaxies, covering the whole sky, and selected using limits in volume, magnitude, and size ($d <40$\,Mpc, $m_B < 15.5, D_{25} > 1$\,arcmin, as obtained from HyperLEDA), though avoiding the plane of the Milky Way ($|b|>30^\circ$).
 
Many of the science themes that can be tackled with S$^4$G data (see Sheth et al. 2010 for an overview, and Mart\'\i n-Navarro et al. 2012 or Elmegreen et al. 2013 for examples) can benefit from, or may even need, additional imaging at optical wavelengths. This can be for various reasons, for instance to compare directly with previous work that will mainly have been done in the optical, to trace different stellar populations, to use mid-IR-optical colour maps or profiles, and/or to explore the effects of dust extinction. In addition, optical data are vital for any analysis of spectral energy distributions in galaxies, which are especially powerful if data at a wider wavelength range can also be incorporated (e.g., from {\it GALEX}). This is illustrated by our recent study of the edge-on galaxy NGC~7241 and its newly discovered foreground companion (Leaman et al. 2014). 

For most of the S$^4$G sample galaxies, optical imaging is publicly available from the Sloan Digital Sky Survey (SDSS; York et al. 2002). In this paper, we present re-processed SDSS images in five bands of 1657 S$^4$G galaxies, which we make available as easy-to-use FITS files that can be downloaded by any interested researcher. These images cover an area of at least 1.5 times the diameter of a galaxy. We add to this set $g$-band images of an additional 111 S$^4$G galaxies which we obtained with the Liverpool Telescope (LT).

In the second part of the current paper, we use the set of newly reduced SDSS images and an extensive database search to identify nearby companions to S$^4$G galaxies, as well as those S$^4$G galaxies that are either interacting or merging. We provide lists and images of these galaxies and their companions for further scrutiny, and derive fractions of interacting galaxies and those with close companions in the local Universe. As the currently favoured cosmological model describing the early and subsequent evolution of structure in the Universe relies on galaxy-galaxy interactions and mergers to explain the current population of galaxies and its properties, the detailed study of interacting galaxies is of paramount importance. Our catalogue of interacting or merging galaxies and of those with companions provides a new sample for such study, while the fractions of such galaxies in the nearby Universe provide important calibrations for any cosmological model incorporating galaxy-galaxy interactions.


\section{Sample selection, observations, and data reduction}

\subsection{Sample selection}

\begin{table*}
\centering
\caption{Details on optical images}
\begin{tabular}{lrcccc}
Galaxy & PGC & Source & Bands & FOV [arcmin] & Calibration\\
\hline\hline
    UGC 12893 &           38 &   SDSS/DR8 &   $u,g,r,i,z$ &    5.0 &   -  \\
     NGC 7814 &          218 &   SDSS/DR7 &   $u,g,r,i,z$ &   13.1 &   -  \\
    UGC 00017 &          255 &   SDSS/DR7 &   $u,g,r,i,z$ &    7.3 &   -  \\
     NGC 7817 &          279 &   SDSS/DR8 &   $u,g,r,i,z$ &    9.9 &   -  \\
     NGC 0014 &          647 &   SDSS/DR7 &   $u,g,r,i,z$ &    5.0 &   -  \\
    UGC 00099 &          757 &   SDSS/DR8 &   $u,g,r,i,z$ &    7.5 &   -  \\
    UGC 00122 &          889 &  LT/RATCam &          $g'$ &    4.6 &     1\\
    UGC 00132 &          924 &   SDSS/DR8 &   $u,g,r,i,z$ &    5.0 &   -  \\
    UGC 00156 &         1107 &   SDSS/DR8 &   $u,g,r,i,z$ &    5.0 &   -  \\
     NGC 0063 &         1160 &   SDSS/DR8 &   $u,g,r,i,z$ &    6.0 &   -  \\
\hline\hline
\end{tabular}
\tablefoot{Sample of the table accompanying the data release on the CDS. Only the first ten lines of the file are shown here. Columns 1 and 2 are the galaxy common name and number from the Catalogue of Principal Galaxies (PGC); column 3 is the origin of the image, identifying the data release in the case of the SDSS, and the instrument in the case of the new LT images; columns 4 and 5 are the bands made available and the size of the images (all images are square); and column 6 identifies the calibration method used in the case of the LT images (see Sect.~2.3).}
\label{CDStable}
\end{table*}

We started out by selecting all those galaxies in the original S$^4$G sample (as defined by Sheth et al. 2010) that had imaging in the seventh Data Release (DR7; Abazajian et al. 2009) of the SDSS. This search yielded 1252 galaxies. Later, we added another 183 galaxies for which imaging was released in SDSS DR8 (Aihara et al. 2011) but which were not in DR7, and 222 galaxies from the new extended S$^4$G for all but five of which we used DR8 data. The total number of galaxies for which we re-processed and now release SDSS imaging is thus 1657. These galaxies are identified in Table~\ref{CDStable}.

We also identified those galaxies in the northern hemisphere (defined here as having declination $>-10\deg$) which were not included in the SDSS imaging survey. This yielded 185 galaxies, of which 10 have diameters $D_{25}$ larger than 8\,arcmin. A total of 111 of these galaxies with diameters smaller than 8\,arcmin (see Table~\ref{CDStable}) was observed by us in only the $g$-band, using the LT on La Palma, as described below.
 
We thus present images in the full set of five SDSS bands ($ugriz$) for 1657 galaxies, and images in the $g'$-band for an additional 111 galaxies. This amounts to optical imaging for 1768 galaxies in total, or two-thirds of the S$^4$G sample. 

\subsection{Sloan Digital Sky Survey imaging}

We use imaging from the SDSS DR7 (for 1257 galaxies) and SDSS-III DR8 (for 400 galaxies). No further imaging data has been, or will be, released by SDSS-III since DR8. The SDSS/SDSS-III data are freely available from the corresponding archive servers. We worked with the ``corrected frames'' produced by the SDSS photometric pipeline, but we carried out our own background subtraction for both datasets.  We then produced FITS files of a size which covers at least three times the diameter of the galaxies but with a minimum size of 5.0\,arcmin, with subtracted backgrounds, and with the images in the five bands aligned so they can be downloaded and used easily. We will describe now some of the main aspects of this process, and of the released images.

\subsubsection{SDSS DR7 mosaics, calibration and image quality}

The SDSS Data Archive Server (DAS) distributes flat-fielded, calibrated images, so-called ``corrected'' (``fpC") frames, taken by the SDSS CCD camera (Gunn et al. 1998). \footnote{The dark current is negligible, the gain is around 3\,e-/ADU, and the readout noise is generally less than or around 5\,e-/pixel. The exact values as measured during the observations are reported in the corresponding ``tsField'' files, although we did not save this specific information in the headers of our mosaics.} Each single frame covers $13.51\times9.83$\,arcmin. As we aim to cover a field of at least three times $D_{25}$, in several cases the mosaics contain two or even more adjacent fields. The sky subtraction was carried out on each single frame prior to the mosaic assembly. 

Before estimating the background level, we subtract the 1000 counts of the {\sc softbias} added to all SDSS images. We apply conservative masking of our target object. The mask has a radius of $1.5\times R_{25}$; for highly inclined objects the mask shape is an ellipse with ellipticity and position angle taken from the HyperLeda Catalog and corresponding to those of the target galaxy. We then measure the fluxes in about ten thousand randomly placed five pixel-wide apertures. We apply a resistant mean to the distribution of the aperture fluxes, and carry out several iterations to mitigate the effect of stars and other background objects. The SDSS background is very smooth: the frame-to-frame variation of the skylevel is usually $\lesssim0.2$\,ADU, provided no large-scale gradient is present. The mean of the bias-free distribution provides a good estimate of the sky background, and is then subtracted from the images. This value is also displayed in the header for every single field contained within the mosaic. After removing the background level, we assemble the final mosaics in all five bands, centred on the target galaxy, by using SWarp (Bertin 2010). 

As the DR7 database covers one single epoch for the entire survey area, the depth of the mosaics corresponds to that of one single exposure in SDSS. We did not include the multi-observations of the SDSS Supernova Legacy Survey for any galaxies that might be located in the survey area of Stripe82 (e.g., Bakos \& Trujillo 2012). The 95\% completeness limits of the images for point sources are 22.0\,mag in $u'$, 22.2 in $g'$, 22.2 in $r'$, 21.3 in $i'$, and 20.5 in $z'$ (Stoughton et al. 2002 [Early Data Release]; Abazajian et al. 2004, 2009 [DR2,DR7]). Because of the exquisite sky background behaviour, the SDSS images can be used to derive surface brightness profiles down to $\mu_{r'} \sim27$\,mag\,arcsec$^{-2}$ (Pohlen \& Trujillo 2006).

The counts in the images are in ADUs. In the header of each image, we provide the magnitude zeropoints (``magzpt'') to convert the number counts $F$ into magnitudes, as ${\rm mag} = -2.5\times\log F+{\rm mag}_{\rm zp}$.

The SDSS DR7 photometric calibration is reliable on a level of  2\%. We opted to follow the traditional recipe to determine the zeropoint of the galaxy mosaic by using the aa, kk, and airmass parameters (the photometric zeropoint, the extinction coefficient and the airmass coefficient). These parameters are provided in the tsField table associated with the exposure containing the target object. The magnitude zeropoint can be given as  $-2.5\times(0.4\times[{\rm aa}+{\rm kk}\times{\rm airmass}]+2.5\times \log(t_{\exp})$, where the exposure time for each pixel equals 53.907456\,s. To calibrate in surface brightness, one uses the pixelscale of 0.396\,arcsec/pixel and the standard expression ${\rm zp}_\mu = {\rm zp}_{\rm mag} + 2.5\times\log ({\rm pixel scale}^2)$.

\subsubsection{DR8 specific issues}

The SDSS-III imaging is available from the SDSS-III Science Archive Server. Starting with DR8, the SDSS data releases distribute flat-fielded, sky-subtracted, and calibrated fields in the form of multi-extension FITS files. The sky subtraction was done by an improved algorithm compared to previous data releases that had generally overestimated the sky level. After masking the brightest sources, the background has been modelled by a spline fit over heavily binned and smoothed data (further details can be found in Blanton et al. 2011). This background model can easily be recovered from the multi-extension FITS files and added back into the images.

To be consistent with the treatment of DR7 imaging, we reproduced the non-sky-subtracted images of the DR8 galaxies. Then, we followed the same steps in creating the mosaics as we did with the DR7 data. Comparison of the DR8 sky level with that derived by us shows that the values agree very well. There is a small systematic offset: our sky values are in the median $0.15\% (\pm0.2\%)$ lower than those of DR8 and we also find a slight dependency on the level of the background itself (the relative difference grows to around 0.3\% for the smallest sky values, and is around zero for the largest). The DR8 sky values are thus slightly overestimated compared to what we determined. This difference could be caused by a different sigma-clipping and/or iteration. This relative difference is rather small, but when studying, for instance, the faint outskirts of galaxies, at levels of $\sim 27$\,mag\arcsec$^{-2}$ in the $r'$-band, the difference in sky level could become important.

The calibration of SDSS-III imaging is fundamentally different from that in SDSS-I/II. There was no auxiliary photometric telescope (PT) involved, and the SDSS-III calibration is based on an internal calibration called ``ubercalibration'' (Padmanabhan et al. 2008; Aihara et al. 2009), which basically compares the overlaps between adjacent scans within the survey. This forces DR8 to be on the same zeropoint on average as the DR7 calibration, but it does not use any data from the PT.

Due to this, in SDSS-III the fluxes are expressed in terms of the rather frivolously named ``nanomaggies'', not ADUs. In this description, a ``maggy'' is the flux $f$ of the source relative to the standard source $f_0$, which defines the zeropoint of the magnitude scale. Therefore, a ``nanomaggy'', or nMgy, is $10^{-9}$\,maggy. To relate these quantities to standard magnitudes, an object with flux $F$ given in nMgy has a Pogson magnitude $m = -2.5\times\log F + 22.5$.

The standard source for each SDSS band is close to but not exactly the AB source (3631\,Jy, Fukugita et al. 1996), meaning that a nanomaggy is approximately $3.631\times 10^{-6}$\,Jy. However, the absolute calibration of the SDSS system has some percent-level offsets relative to AB. The $u$ band zeropoint is in error by 0.04\,mag, in the sense that $u_{\rm AB} = u_{\rm SDSS} - 0.04$\,mag, while $g, r$, and $i$ are close to AB. The $z$-band zeropoint is not as certain, and may be shifted by about 0.02\,mag in the sense $z_{\rm AB} = z_{\rm SDSS} + 0.02$\,mag (Abazajian et al. 2004; the same offsets have to be applied to the DR7 imaging as well, when transforming into the AB system.) 

\subsubsection{Advantages of the re-reduced images}

Through this release, the optical imaging for two-thirds of the S$^4$G sample galaxies can easily be accessed in FITS format. The naming convention includes the identification of the target galaxy and the filter, for instance, the $r'$-band mosaic of PGC~01107 is called: PGC01107rmosaic.fits. The sky-subtraction algorithm and the mosaicing were both optimised to cope with the presence of large galaxies. We inspected all images visually, and manually corrected any discrepancies where the automatic pipeline might have failed. The large image size (at least $3\times D_{25}$ is covered in all filters) is an important aspect which allows one to explore (1) the immediate vicinity of these galaxies, and (2) the faintest structures observable with SDSS imaging, including surface brightness profiles out to $\sim2\times R_{25}$. Our images obviously maintain the excellent background and noise characteristics, and excellent photometric and astrometric calibration, of the SDSS image survey. An illustration of these advantages can be seen in Fig.~\ref{N4321}.

\subsection{Liverpool Telescope imaging}

The LT (Steele et al. 2004) is a 2.0\,m robotic telescope situated in the Roque de Los Muchachos Observatory on the island of La Palma. We have used two different cameras for our project, namely RATCam and IO:O, to observe a total of 111 S$^4$G galaxies which had not been observed in the SDSS, are smaller than 8\,arcmin, and at a declination $>-10\deg$.

RATCam has now been retired from the LT. The field of view (FOV) was $4.6\times4.6$\,arcmin, with a pixel scale of 0.1395\,arcsec/pixel. We observed 5 galaxies with RATCam between September and October 2011, using a Sloan $g'$-band filter. Each galaxy was observed for $3\times100$\,s.

RATCam was replaced by the IO:O camera\footnote{http://telescope.livjm.ac.uk/Info/TelInst/Inst/IOO/}. It provides a wider field of view and greater sensitivity than RATCam. The FOV is $10\times 10$\,arcmin and the pixel scale is approximately 0.15\,arcsec/pixel. However, we used the $2\times2$ binning because the $1\times1$ binning was not operative at the time of the observations, with a resulting pixel scale of 0.3008\,arcsec/pixel. We observed 106 galaxies with IO:O using a Sloan $g'$-band filter, also with $3\times100$\,s exposure time, in February 2012 and from July 2012 to January 2013.

LT images are delivered after having passed a basic data reduction pipeline, which includes bias subtraction and flat-fielding correction. The overscan regions are trimmed off,  leaving a $2048\times 2048$ pixel image. We subtracted the background before, and again after the combination of the three separate exposures by determining the background from regions well outside the galaxy, ignoring emission from, e.g., foreground stars or cosmetic defects by combining with a median algorithm. If one of the three individual exposures had significantly different seeing from the others, it was rejected before the combination. Also, if some of the frames presented strange features (e.g., a satellite track, humidity problems), the corresponding images were rejected. For flux calibration, a set of Landolt standard stars was observed in all the bands every two hours, specifically PG0231+051, RUBIN~149, PG1047+003, PG1525-071A, MARK-A and PG2331-055A. To transform from the $UBVR_{\rm cIc}$ filter set used in the literature to $ugriz$, we used the equations presented in Smith et al. (2002). We computed the corresponding zero-points to perform the photometry, and added them to the headers of the images (stored in keyword COMMENT2). For the nights when no standard stars were observed, we performed the flux calibration by computing the $g'$ magnitude using $BVRI$ photometry values obtained from the NED, as described below.

The headers contain a keyword COMMENT3 with information about the calibration method used. This information is expanded in Table~\ref{CDStable}, as follows: 1. -- standard stars, 2. -- cross-calibrated using a total $g'$ magnitude as derived from photometry in NED, from different bands, namely 2.1 -- $g'$ computed from $B$ and $V$, 2.2 -- computed from $B$ and $R$, 2.3 -- from $B$ and $I$, 2.4 -- from $B$, $R$ and $I$, and 2.5 -- from $b$ and $J$. In all cases we performed linear correlations using predictions based on the MIUSCAT stellar populations models. This set of predictions is an extension of the Vazdekis et al. (2003, 2010) models, based on the Indo-U.S., CaT and MILES empirical stellar libraries. A full description of the models is given in Vazdekis et al. (2012) and applications are provided in Ricciardelli et al. (2012). The equations used are $g' =R+0.7232(B-R)+0.0224$, $g' =I+0.7902(B-I)+0.0517$ and $g' =J+0.8413(B-J)+0.0927$, with a correlation coefficient $R^2$ of over 99.5\%. The LT images of four galaxies have a value X in keywords COMMENT2 and COMMENT3. These images, of UGC~09992, UGC~10194, PGC~027825 and PGC~029086, could not be calibrated because we did not have standard star observations, nor photometric information to calculate a total $g'$ magnitude. In spite of this, we do release the images.

The final images have a seeing between 1 and 3\,arcsec, with a median seeing value of 1.4\,arcsec. The final list of galaxies for which we present LT $g'$-band images is shown in Table~\ref{CDStable}, which also lists other basic parameters of the images such as the instrument and calibration method used. 


\section{Image quality control}

We checked all images by eye, aiming to ensure that all images are properly reduced and consistent. The LT images were checked during the data reduction procedure. As there were relatively few LT images and as they were treated individually, no additional quality control was necessary. For the SDSS images, which constitute the bulk of our images (over 8000 in total), we developed two simple scripts, the first of which identifies images with more than a small number of non-physical pixels, and the second of which displays the images in all five bands for a certain galaxy in one image display window, scaling them logarithmically. 

We thus identified a few images with problems, which were corrected by re-processing the data. A small number of galaxies ($<1\%$ of the total) were found to be near the edge of the SDSS survey area. We left these in as the images may still be useful, even if the outer region of a galaxy is not completely  visible (e.g., NGC~6118). Finally, when producing the false-colour images we realised that a small fraction ($\sim2\%$) of the DR7 images had a different astrometric orientation, X=N and $-$Y=E instead of Y=N and $-$X=E. For some reason which we do not understand and which we have not seen documented in the literature, galaxies at low latitudes are sometimes given this incorrect astrometric solution. We easily corrected this by applying a rotation-transposition-rotation transformation. An example of this is NGC~1055 (the various SDSS images shown on the `images' page for this galaxy on the NED illustrate the problem). After these checks, all released images should be directly useful for science, although given the large number of them we cannot guarantee the complete absence of small defects.


  \begin{figure*}
   \centering
   \includegraphics[width=0.95\textwidth]{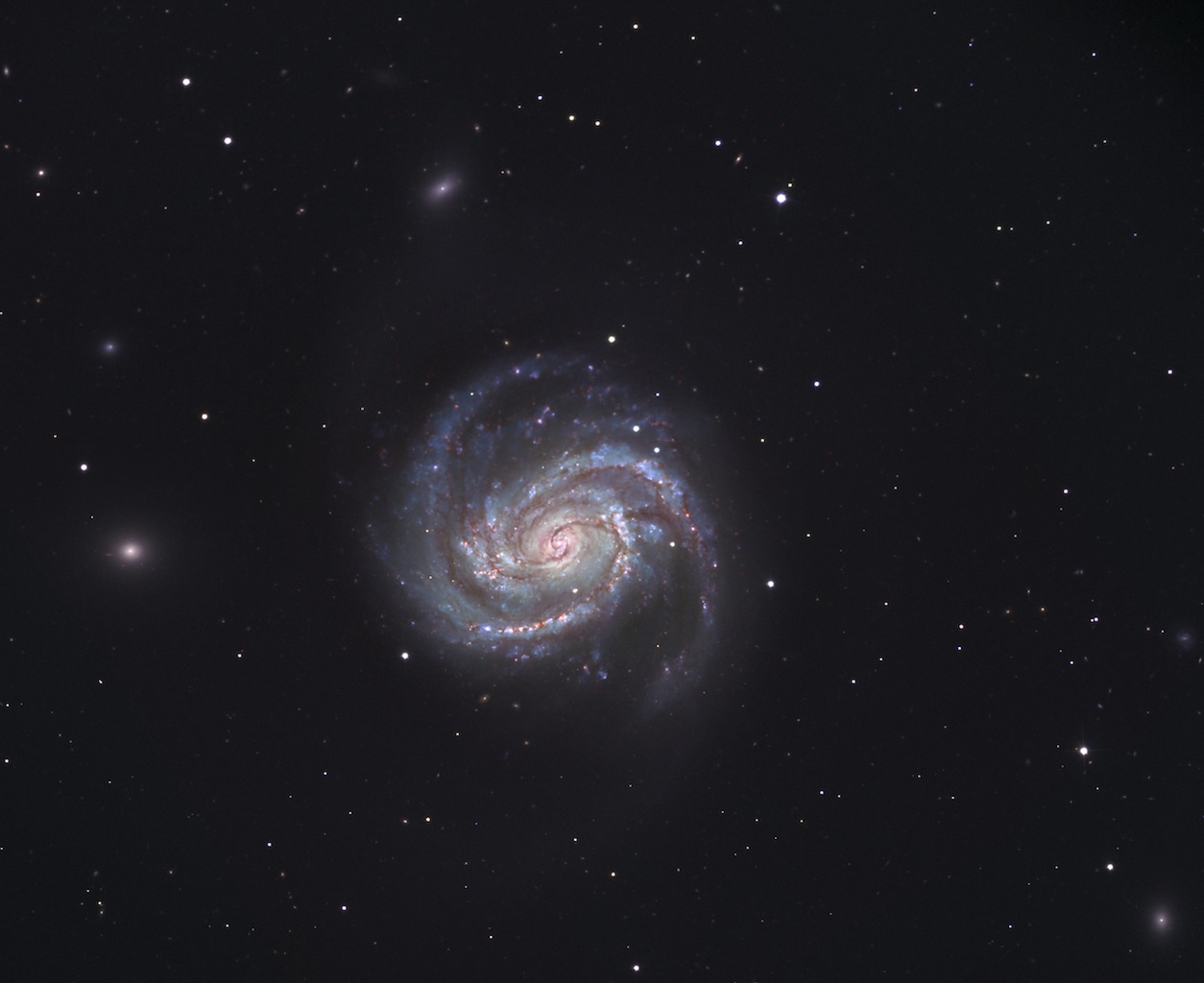}
   \caption{False-colour image of the Virgo cluster spiral NGC~4321 (M100).}
              \label{N4321}%
    \end{figure*}

\section{False-colour images}

We used the images described above to produce false-colour images of 29 selected galaxies from our sample, which we release publicly for their use in, for instance, scientific presentations and in public outreach work. For this, we selected primarily large galaxies, which will yield images with high resolution across the galaxies and their components, and particularly interesting galaxies, such as interacting and merging galaxies, or with features such as rings, loops, warps, or dust lanes. All attempts were made to produce aesthetically pleasing images that attempt to capture appropriate details and colours approximating true-colour images.

To produce these images, an example of which is shown in Fig.~\ref{N4321} for the galaxy NGC~4321, we scale the different images in such a way as to show their structure in the most clear and aesthetically pleasing way, rather than in an alternative way which would be to scale them approximately and then combine them to produce the equivalent of an ``RGB'' image with no further cosmetic work. The latter approach may be more reproducible but will lead to ugly images. 

The first step in the production of these images is to scale the raw FITS files to show the outer spiral arms as well as the detail in the centre of the galaxy. We start out with three images, usually those in the $g, r$ and $i$ bands, and then include the {\it Spitzer} 3.6\,$\mu$m S$^4$G image as obtained from the IRSA archive, and in most cases the SDSS $u$ and $z$-band images, although the latter do not tend to add significantly to the end result. Aiming to represent typical expected colours such as blue for young stellar populations and red and brown for dust lanes, we use the $g$-band data as the blue channel, the $r$-band data for red, and a mixture of $u$, $z$ and 3.6\,$\mu$m added to the $i$-band data for the green channel in order to produce a balanced equivalent to three-band images and to highlight certain features that were enhanced by using in particular the {\it Spitzer} infrared data. This methodology introduces a lot of "false'' colour into the images but is ultimately the most successful way to produce images for the use intended here, which is not so much scientific, but primarily for illustrative and public outreach purposes.

The effort in producing the final images comes mainly from the combination of the various images, their scaling, correction of background levels, and cosmetic edits (including colour shifts, cosmetic repair of uneven sky backgrounds, or shifted star colours). 

\section{Image availability}

\subsection{Science images}

All science images are released publicly with the publication of this paper. They can be used freely by any researcher interested, provided the origin of the images is identified as this paper, SDSS or LT, and the data archive they were downloaded from. Images are available through the NASA-IPAC Extragalactic Database (NED) and the Centre de Donn\'ees astronomiques de Strasbourg\footnote{Via anonymous ftp to cdsarc.u-strasbg.fr (130.79.128.5) or via http://cdsarc.u-strasbg.fr/viz-bin/qcat?J/A+A/xxx/yyy} (CDS) in FITS format. The headers of the images give information on the origin of the image, filter and other instrumental parameters, and calibration information. A README file accompanies the data release on the CDS, listing the galaxies, the images with their sources, the size of the images, and the calibration method used in the case of LT data. The first ten lines of this README file are shown here as Table~\ref{CDStable}. 

The S$^4$G images of these galaxies in the 3.6 and 4.5\,$\mu$m bands are released by the S$^4$G team through the IRSA, and do not form part of the data release related to this paper.

\subsection{False-colour images}

The resulting false-colour images are available publicly from the website of the EU-funded Initial Training Network DAGAL (Detailed Anatomy of GALaxies, www.dagalnetwork.eu), and can be used freely for any purpose provided the origin of the images is clearly acknowledged. We supply the images as jpeg files of various sizes, as well as high-resolution tiff files for use in professional printing applications. 


\section{Interacting and companion galaxies}

We used our visual inspection of all images of over 1700 nearby galaxies with SDSS imaging as the starting point to make an inventory of which of the S$^4$G sample galaxies have a close companion, and which are interacting or merging. This allows a direct estimate of the fraction of galaxies in these categories, and defines samples for future study, either of individual galaxies, or of their collective properties, such as for example the star formation rate (which has been found to be enhanced in galaxies with a close companion, see, e.g., Knapen \& James 2009 and references therein).

\subsection{Close companions}

To define which galaxies have a close companion nearby and massive enough to plausibly cause gravitational tidal effects, we follow the approach of Knapen \& James (2009), which in turn was based on works by Schmitt (2001), Laine et al. (2002), and Knapen (2005). We consider a galaxy to have a close companion if this companion (1) is within a radius (measured from the centre of the galaxy) of five times the diameter of the sample galaxy, or $r_{\rm comp} < 5 \times D_{25}$, with $D_{25}$ from the RC3, {\it and} (2) has a recession velocity within a range of $\pm200$\,km\,s$^{-1}$ of the galaxy under consideration, {\it and} (3) is not more than 3 mag fainter (using magnitudes in NED). As a rough indication, $5 \times D_{25}$ corresponds to around 50\,kpc in most survey galaxies, although in some cases this distance can be as large as 200\,kpc (depending on the size of a galaxy and its distance to us). The precise value for each galaxy with a close companion can be derived from Table~3.

To find galaxies that qualify under these criteria, we performed an automated search of all 2829 galaxies in the S$^4$G sample (original plus extended) using NED. We then filtered the results to end up with a list of those galaxies that have a close companion, and the name and basic properties of the companion. We went through the list to remove any erroneous entries, such as duplicate galaxies (where NED has the same object listed under different names, with sometimes slightly different positions), or sources within galaxies (e.g., X-ray sources, H{\sc ii} regions).

The list of galaxies with {\it bona fide} companions is given in Table~3. A total of 470 galaxies of our sample of 2829 galaxies have a companion by the criteria identified above (17\%). One can argue whether these criteria are optimal, but we feel that they are a good compromise to find companions which are massive and nearby enough so they can be expected to have a tidal gravitational impact on the sample galaxy. Our second criterion, of relative recession velocities within $\pm200\,{\rm km\,s}^{-1}$, can be widened, but more unrelated neighbouring galaxies would creep in. However, we do note the case of NGC~3384 and its companion M\,105 (NGC~3379) at 7.2 arcmin, 0.6\, mag brighter, and $\Delta v = 207\,{\rm km\,s}^{-1}$, where we are most likely dealing with a galaxy pair, but which cannot be formally catalogued here as a close companion. In addition, two galaxies which were catalogued as having a close companion have other companions which fall just outside the criteria (NGC~4106 with companion NGC~4105 at 1\,arcmin, 0.76\,mag brighter, and $\Delta v = -214\,{\rm km\,s}^{-1}$; and NGC~5560 with companion NGC~5566 at 5.2\,arcmin, 1.7\,mag brighter, and $\Delta v = -222\,{\rm km\,s}^{-1}$). As we only identified these cases, we estimate that such effects introduce an uncertainty of at most a percent.

  \begin{figure*}
   \centering
 \includegraphics[width=0.75\textwidth]{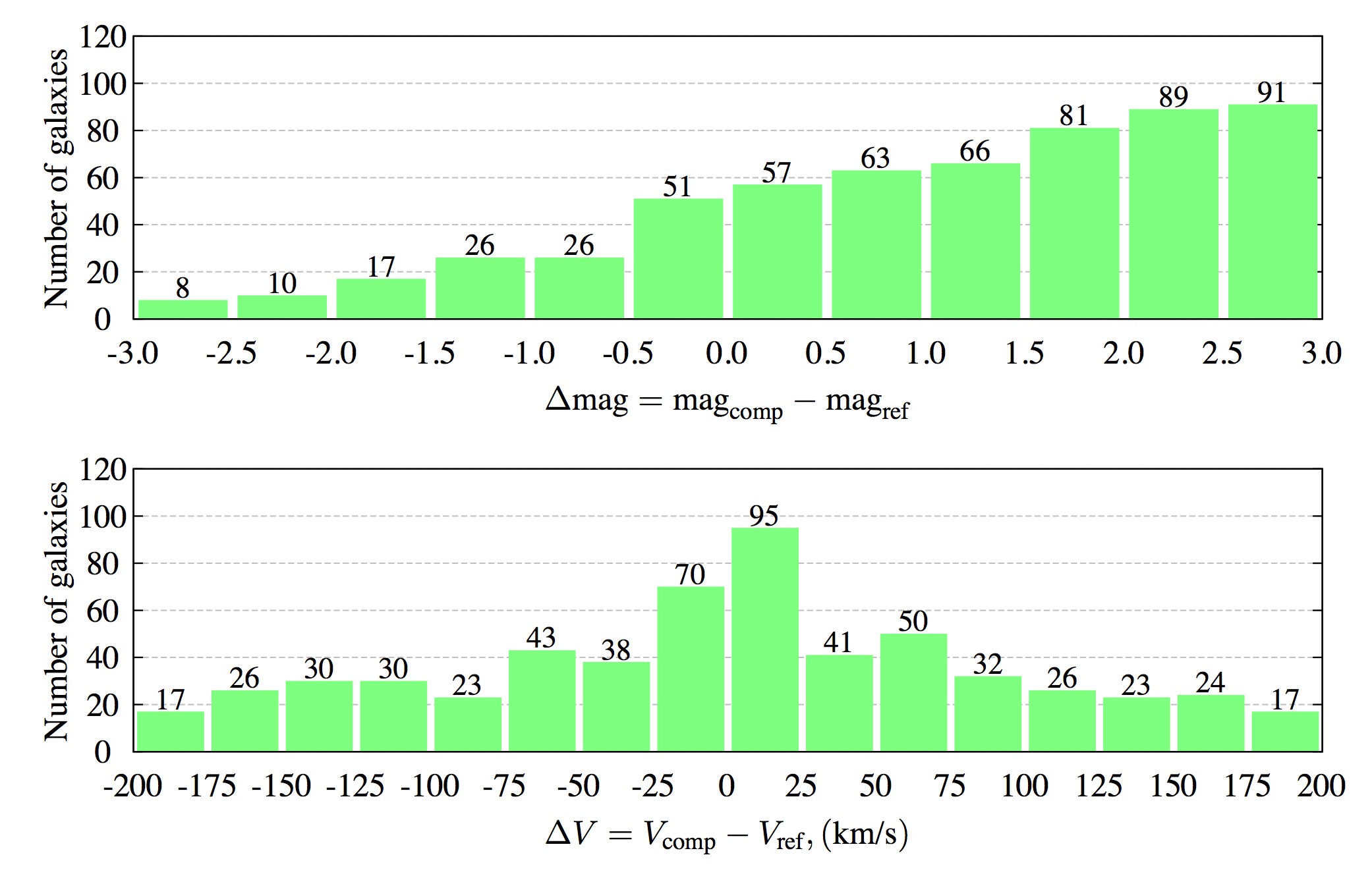}
   \caption{Histograms showing the distribution of magnitude (top) and recession velocity difference (lower panel)  for galaxy-companion pairs.}
              \label{histo}%
    \end{figure*}
  
It is possible that companions fainter than 3\,mag are not catalogued in NED, in particular this is likely to be the case for the faintest of our sample galaxies. We thus calculate the fraction of bright (<13.5\,mag, we use this limit because the NED should be complete down to 16.5\,mag) galaxies with a close companion, which is  21\% (267 of 1281 galaxies), and confirm that this is the case. Figure~\ref{histo} illustrates how the magnitude difference criterion introduces an asymmetry in the results: many more fainter companions are found than brighter ones. For comparison, the lower panel of Fig.~\ref{histo} shows that the velocity difference between a sample galaxy and its companion is indeed symmetric around 0\,km\,s$^{-1}$, as expected. Note that we considered every S$^4$G sample galaxy separately, and a pair of companions of which both members are in the S$^4$G sample will thus be counted twice. Also note that companions which are themselves not  S$^4$G sample galaxies are included in this plot, so the total number of galaxies included here is larger than the number of S$^4$G galaxies with companions.

\subsection{Interacting and merging galaxies}

  \begin{figure*}
   \centering
   \includegraphics[width=0.95\textwidth]{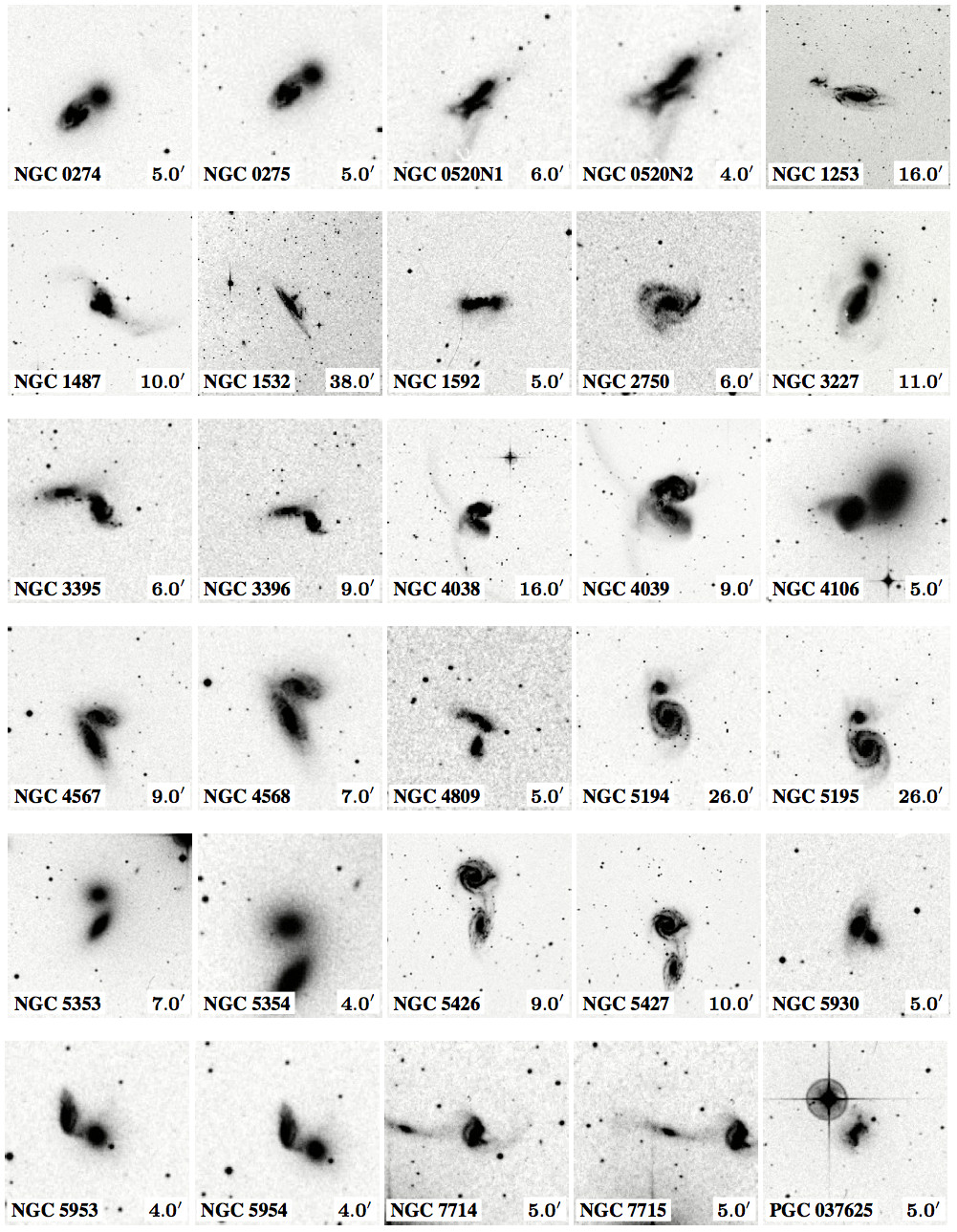}
   \caption{Class A galaxies, similar-size galaxies which are overlapping and very clearly interacting. Details of the S$^4$G sample galaxies and the companions are in Table~3. Images are centred on the galaxy which is identified in the bottom-left label, while the bottom-right label indicates the size, in minutes of arc, of the image shown.}
              \label{ClassA}%
    \end{figure*}

\setcounter{figure}{2}   

  \begin{figure*}
   \centering
      \includegraphics[width=0.95\textwidth]{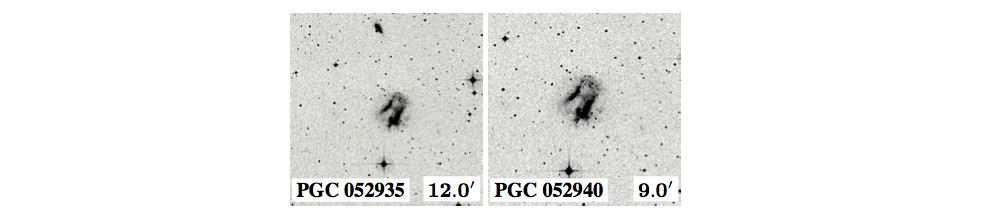}
   \caption{Continued.}
    \end{figure*}

  \begin{figure*}
   \centering
   \includegraphics[width=0.95\textwidth]{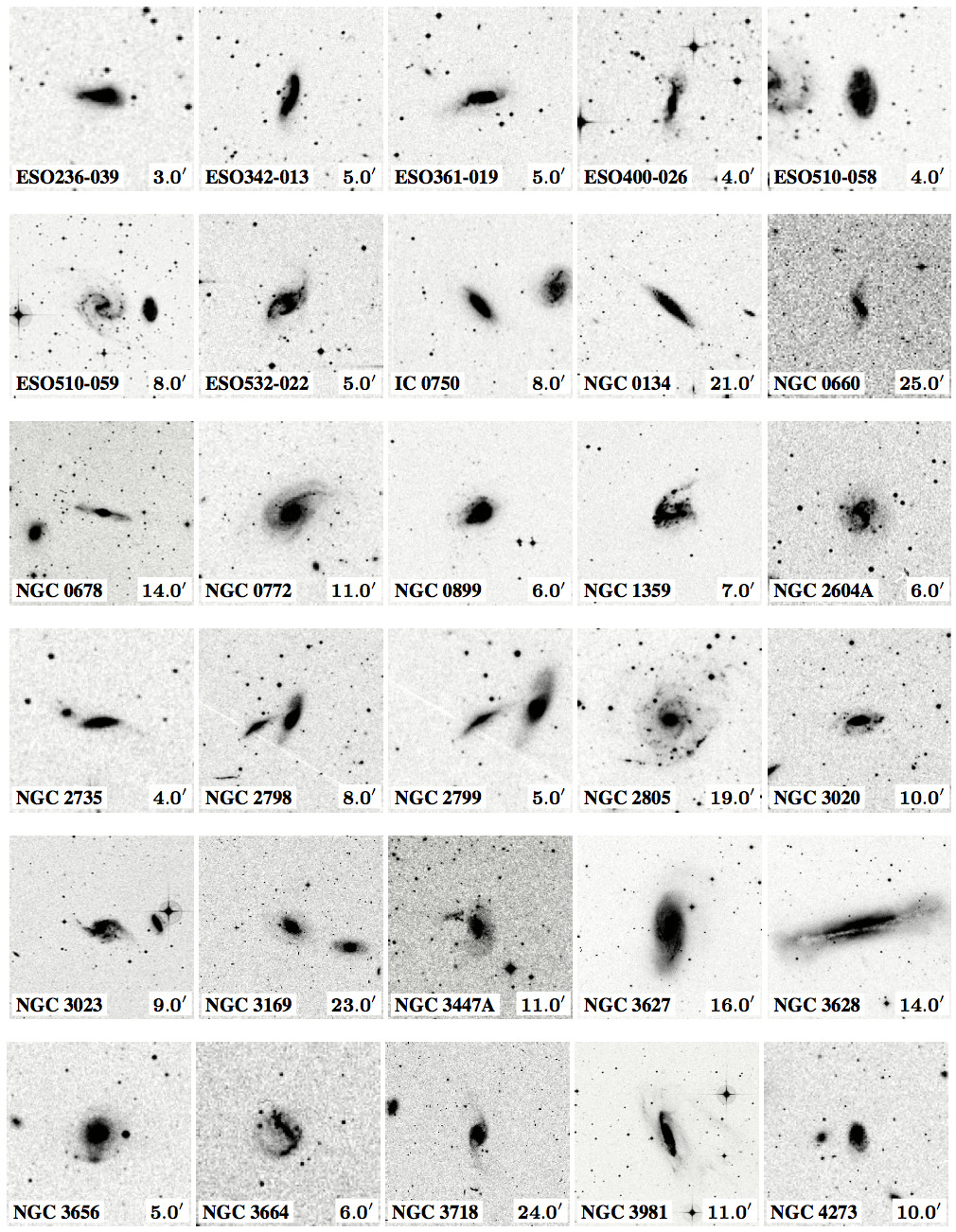}
   \caption{As Fig.\ref{ClassA}, now for Class B galaxies, showing significant signs of interaction.}
              \label{ClassB}%
    \end{figure*}
 
 \setcounter{figure}{3}   

  \begin{figure*}
   \centering
      \includegraphics[width=0.95\textwidth]{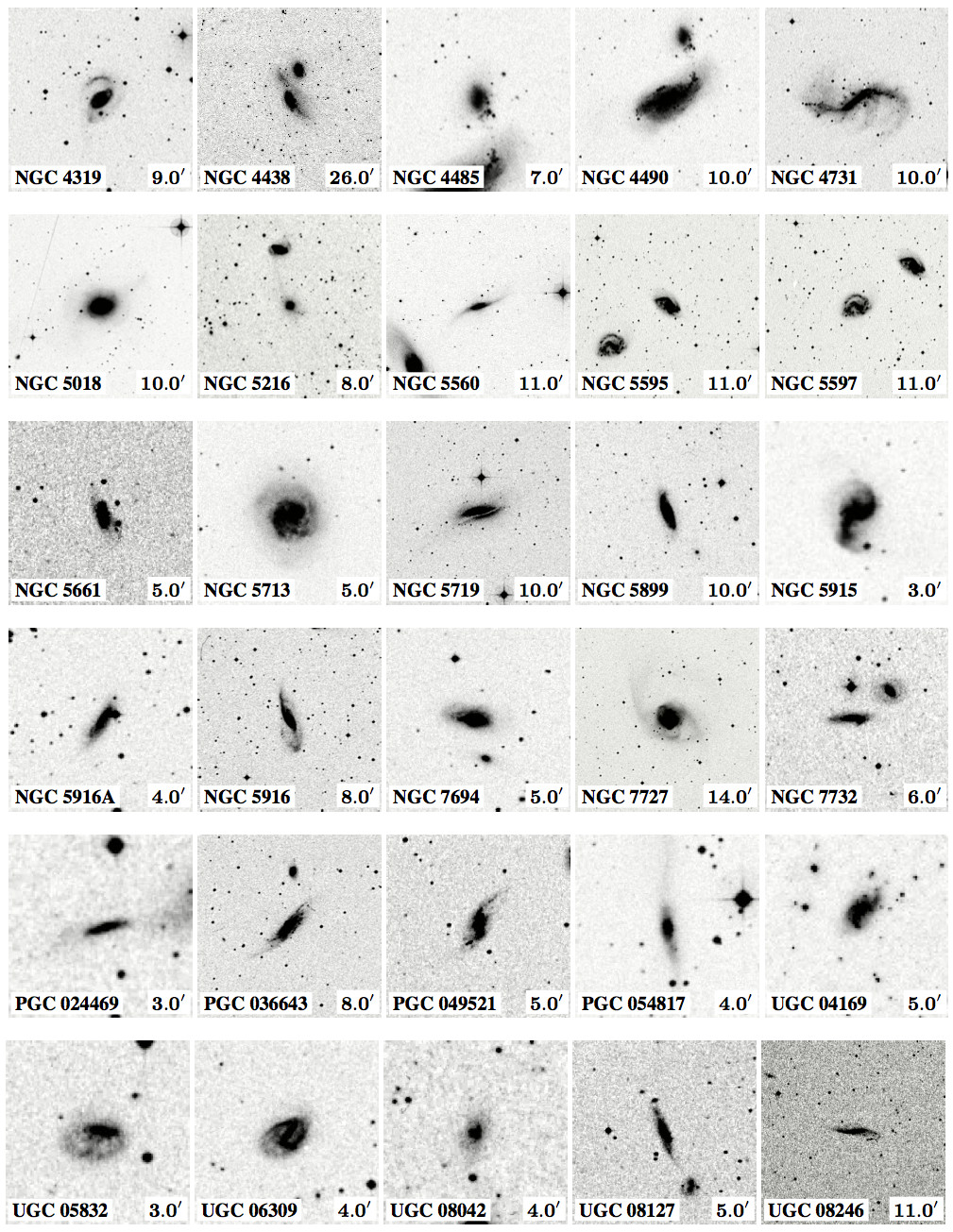}
   \caption{Continued.}
    \end{figure*}

 \setcounter{figure}{3}   

  \begin{figure*}
   \centering
      \includegraphics[width=0.95\textwidth]{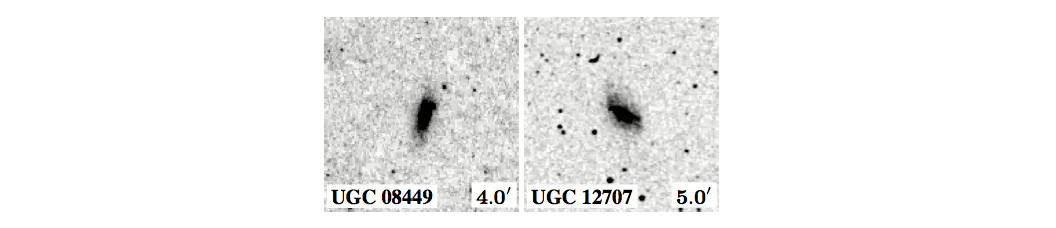}
   \caption{Continued.}
    \end{figure*}

  \begin{figure*}
   \centering
   \includegraphics[width=0.95\textwidth]{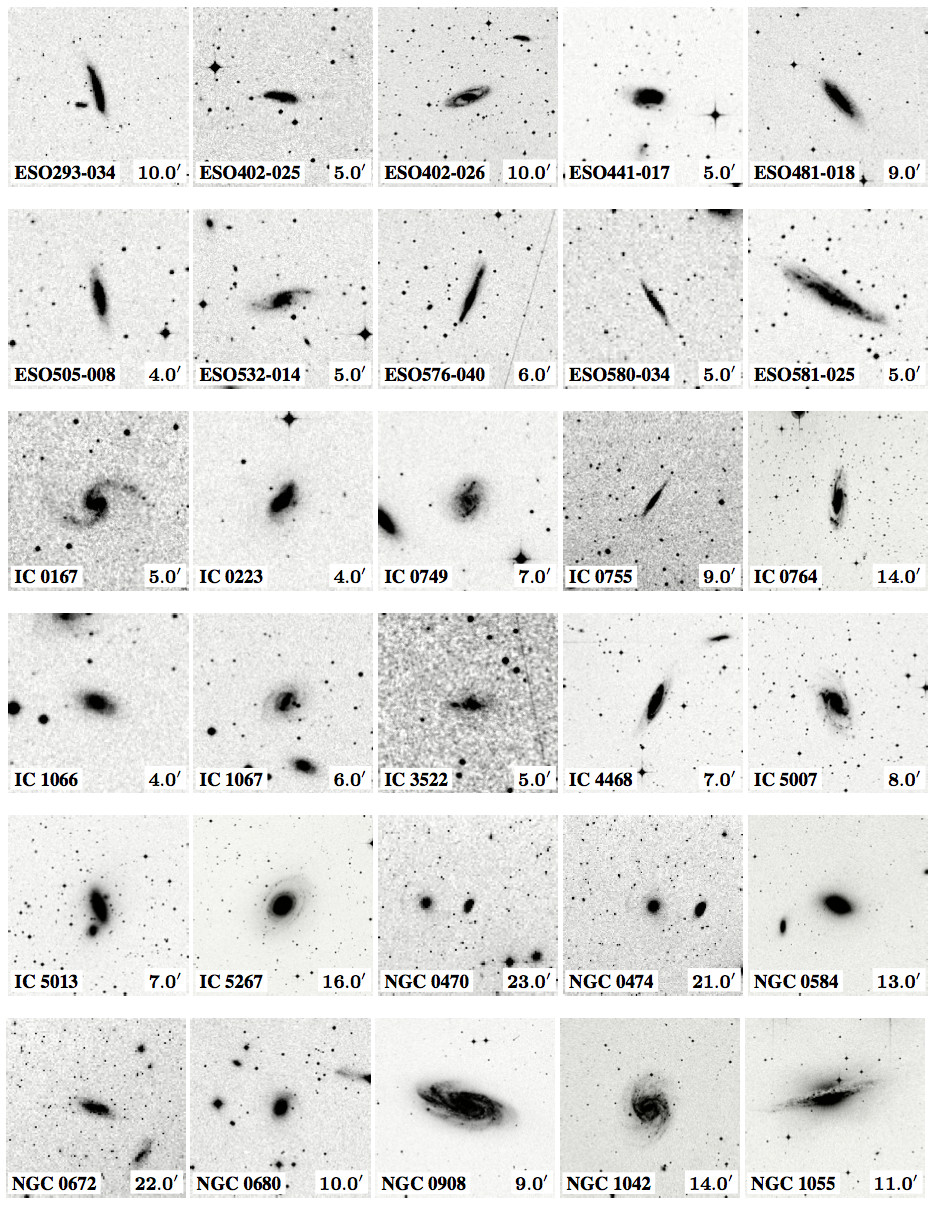}
   \caption{As Fig.\ref{ClassA}, now for Class C galaxies, showing some sign of interaction. Not all these signs may be visible on these small images, and at this particular choice of display levels.}
              \label{ClassC}%
    \end{figure*}
 
 \setcounter{figure}{4}   
  \begin{figure*}
   \centering
   \includegraphics[width=0.95\textwidth]{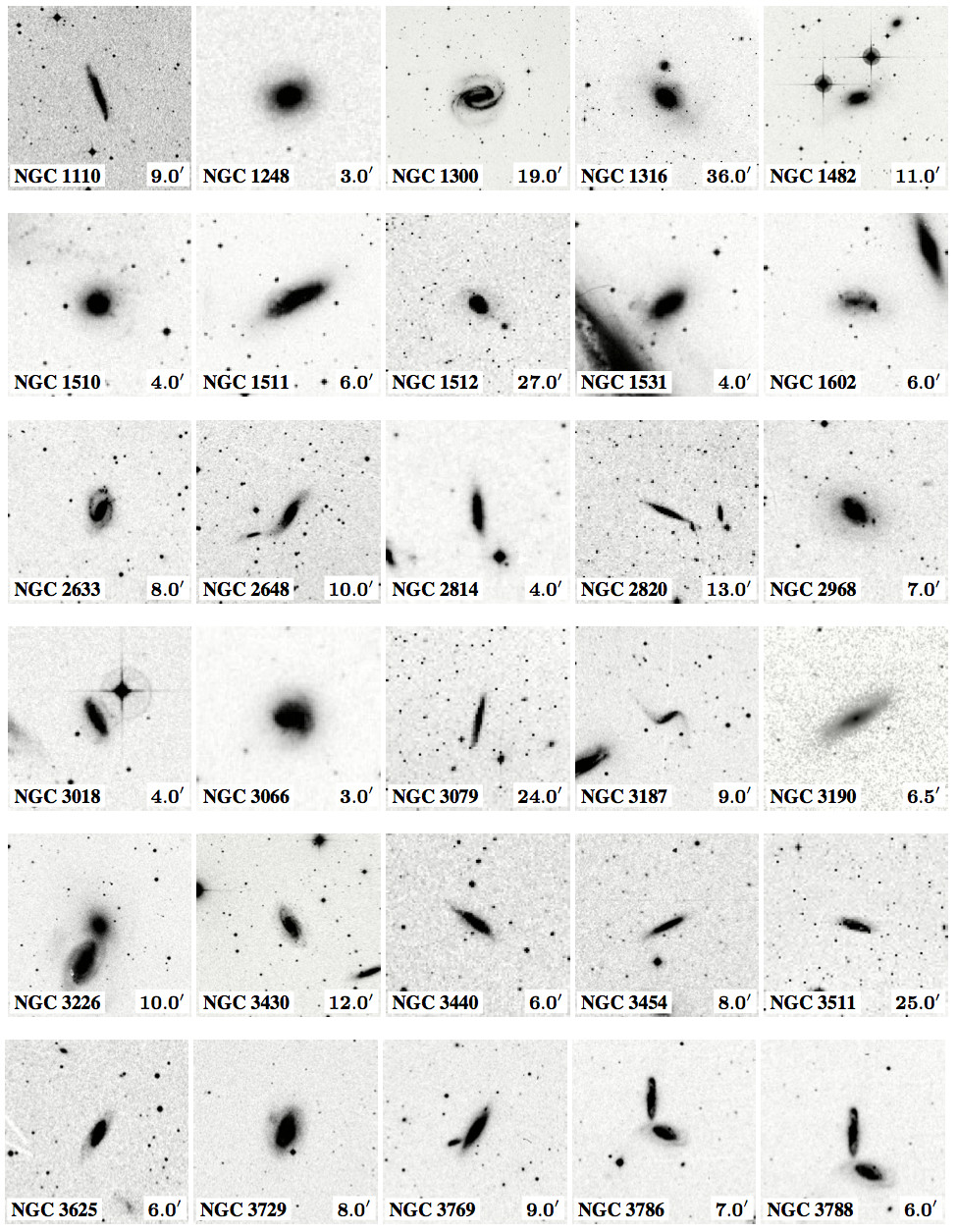}
   \caption{Continued.}
    \end{figure*}
    
  \setcounter{figure}{4}   
  \begin{figure*}
   \centering
   \includegraphics[width=0.95\textwidth]{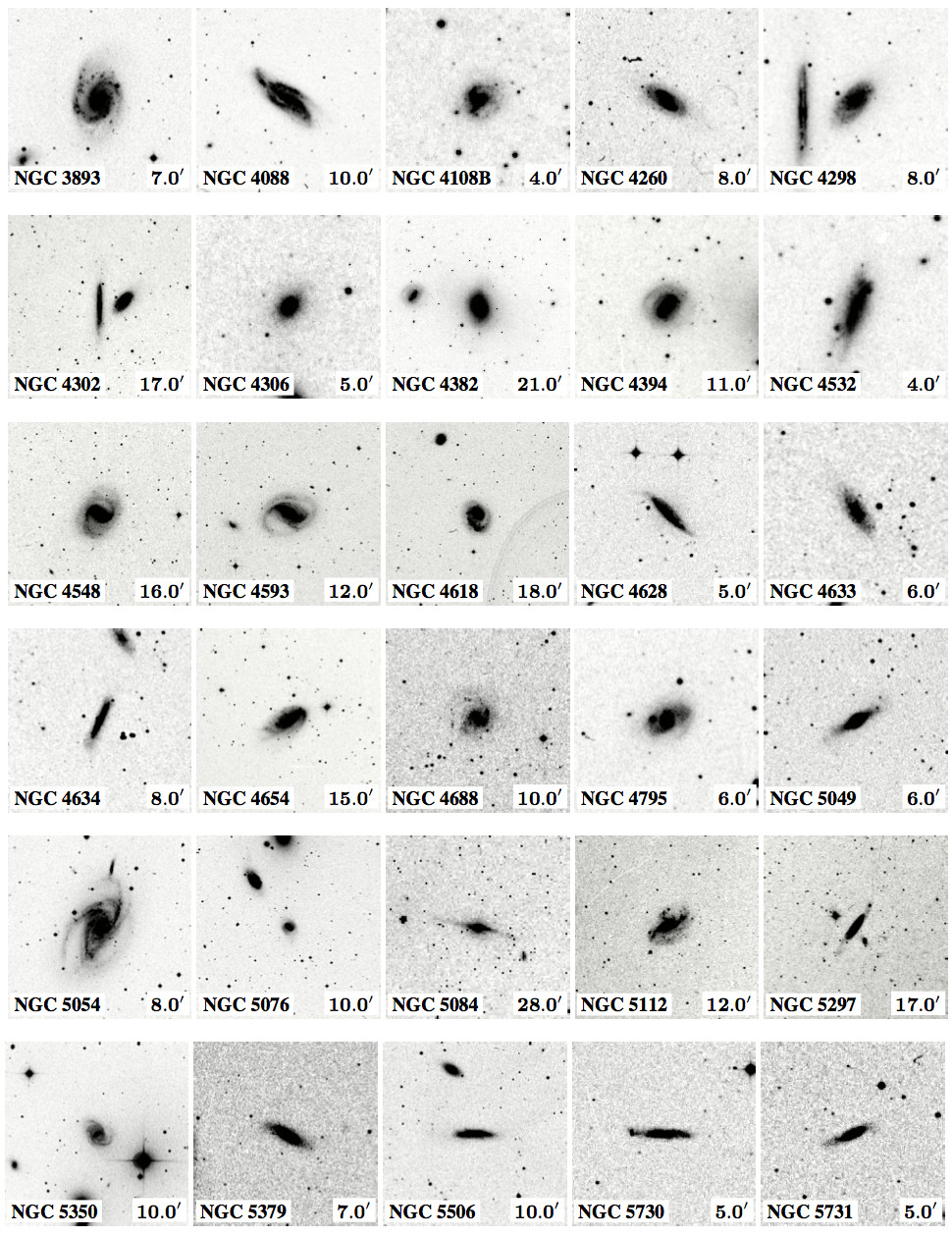}
   \caption{Continued.}
    \end{figure*}
 
  \setcounter{figure}{4}   
  \begin{figure*}
   \centering
   \includegraphics[width=0.95\textwidth]{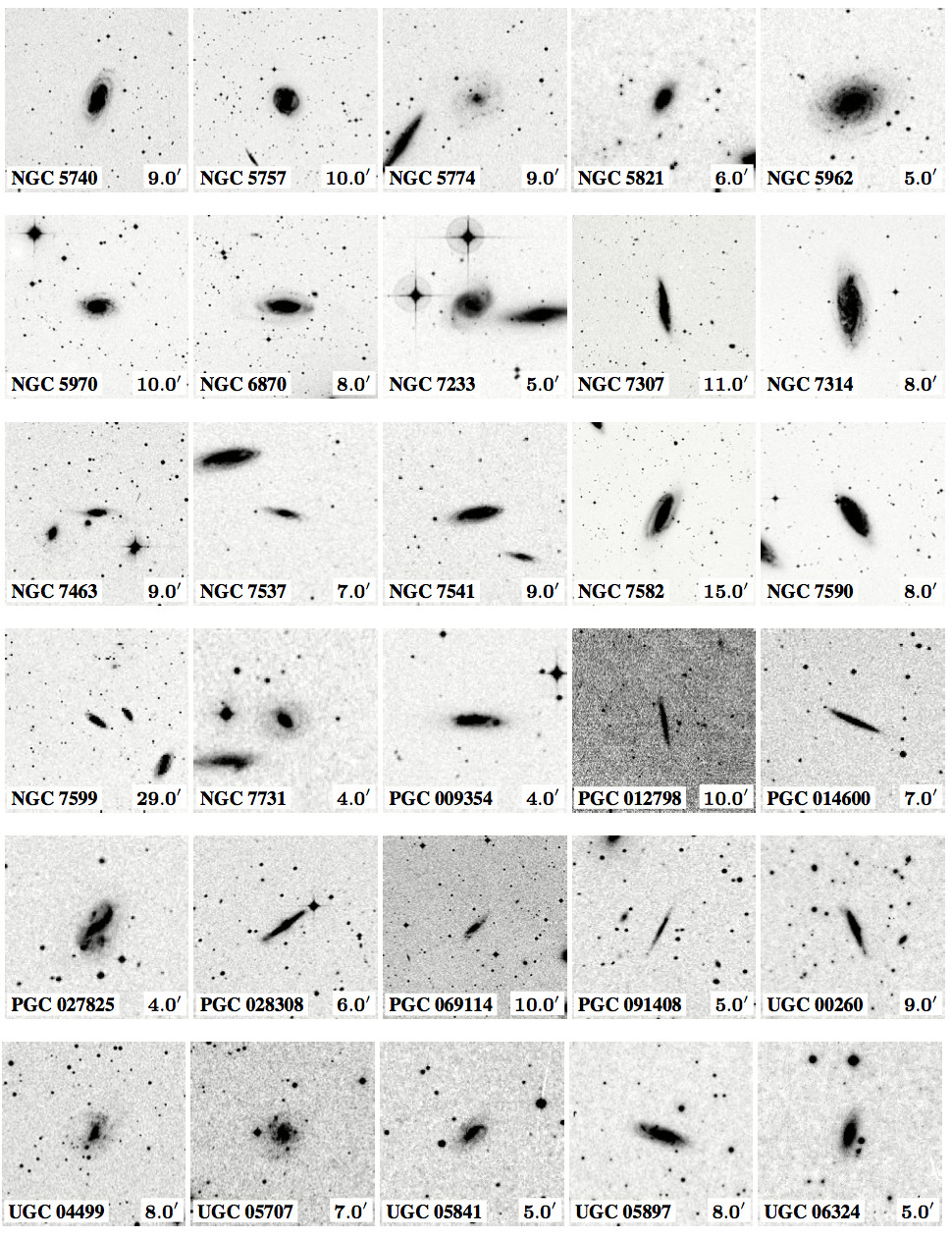}
   \caption{Continued.}
    \end{figure*}
 
    \setcounter{figure}{4}   
  \begin{figure*}
   \centering
   \includegraphics[width=0.95\textwidth]{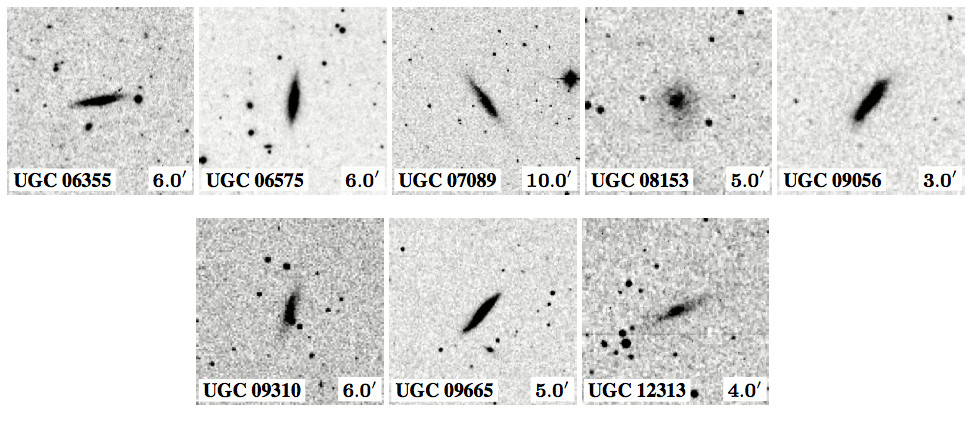}
   \caption{Continued.}
    \end{figure*}
   
In the next step of our analysis, we visually study all galaxies identified to have a close companion in Sect.~6.1, and classify them by the degree of disturbance in appearance related to tidal interactions with the companion. We classify galaxies in three categories for this purpose, plus the remaining and largest category of galaxies which appear undisturbed. The three categories are:

A. Mergers: two similar-size galaxies which are overlapping and very clearly interacting;

B. Highly distorted galaxies, which are interacting as can be judged by morphological hallmarks as tidal arms, or gross distortions of the stellar disk; and

C. Galaxies which show some sign of interaction, such as relatively minor distortions of their disk, or minor tidal features.

As this is based on visual classifications, the criteria are somewhat subjective, but this exercise is nevertheless very efficient to classify galaxies at various stages of interaction and merging. In total, we find 32 (1.1\% of the complete S$^4$G sample and 7\% of the 470 galaxies with close companions) Class-A mergers,  62 (2.2\% and 13\%) Class-B highly distorted interacting galaxies, and 128 (4.6\% and 27\%) Class-C interacting galaxies with minor distortions. The remaining galaxies, with a close companion but not in categories A, B, or C, are 248 in total, or 53\% of the close companion galaxies. Images for all Class A, B, and C galaxies are shown in Figs.~\ref{ClassA}, \ref{ClassB} and \ref{ClassC}, and these galaxies are identified as such in Table~3.

\subsubsection{Comparison with other studies}

The Class A, B, and C used here are very similar to the distortion or perturbation indicator (from 0 to 1 on a scale of 5) of the galaxies in the EFIGI catalogue by Baillard et al. (2011), to the categories $M$ (merging, roughly our Class A), $T$ (tidal, comparable to our B and C) and $N$ (non-disturbed) of Lambas et al. (2012), or to the  morphological perturbation classes $P=0-4$ as defined by Nazaryan et al. (2014). In fact, Lambas et al. and Nazaryan et al. report similar percentages as we do here, namely 60\% for $N$, 30\% for $T$ and 10\% for $M$ (Lambas et al. 2012), and 45\%, 23\%, 22\% and 10\% for classes $P$ 0 to 3, respectively (Nazaryan et al. 2014), corresponding to our measured fractions of 53\%, 27\%, 13\%, and 7\% (see previous subsection) for galaxies with a close companion but without signs of interaction, and Class C, B, and A, respectively. The differences can easily be explained by the quite different selection criteria to select close pairs. In particular the fraction of merging galaxies is stable at around 10\% among pairs of galaxies in the three studies compared. Other studies using sometimes very different methods yield very similar percentages (e.g., Patton et al. 1997, Darg et al. 2009, Holwerda et al. 2011).

We have explicitly considered the overlap in samples between the EFIGI catalogue and ours, to see how well the perturbation indicator of Bailard et al. (2011) correlates with our interaction classes. There are 199 galaxies in common with those of our galaxies which have a close companion (8 in class A, 27 in class B, 47 in class C, and 117 without interaction class). The results of the comparison show that, considering the differences in approach, definitions and data quality, these studies are compatible (the mean perturbation index of our Class A galaxies is $0.67\pm0.07$, for B $0.62\pm0.05$, C $0.34\pm0.04$, and for our remaining galaxies $0.13\pm0.02$. 

The S$^4$G team have recently performed a complimentary study by carefully studying the very outskirts of the S$^4$G galaxies (original sample only) to catalogue any deviations from symmetry at the lowest light levels (Laine et al. 2014). Among other characteristics such as warps, rings, and asymmetries, Laine et al. catalogue from inspection by eye those galaxies with morphological evidence of interactions and past or ongoing mergers. In this, Laine et al. limit themselves to companions visible in the S$^4$G image, use a permitted velocity range of 600\,km\,s$^{-1}$ between sample galaxy and companion, and do not reject all cases where a velocity confirmation cannot be found. Even though the approaches are very different, we find a satisfactory degree of complimentarity, with 64\% of the galaxies found to have a companion in Laine et al. being classified as having one in the current paper, and 69\% for the interacting galaxies.  

These various comparisons with other works in the literature show that an analysis such as the one presented here gives valid and reproducible overall results.

\subsubsection{Interacting fraction among bright galaxies}

We observe that the fraction of interacting galaxies among the PGC, IC, ESO and UGC galaxies in the S$^4$G sample is much lower than among the NGC galaxies. The median magnitudes of NGC, IC, ESO, UGC, and PGC galaxies are 12.81, 14.00, 14.66, 14.89, and 14.79, respectively. As the NGC galaxies are brighter and we noted before the possible incompleteness of faint companions to faint sample galaxies, we consider the fraction of interacting galaxies among the NGC sub-sample as a more accurate estimate for the nearby Universe. We thus find that $350\pm16$ of the 1472 NGC galaxies in the S$^4$G sample have a close companion, or $23.7\%\pm1.1\%$ (where all uncertainties quoted are Poisson errors, cf. Laine et al. 2002). 29 (2.0\%) are Class A galaxies, 42 (2.9\%) are Class B, and 85 (5.8\%) are Class C. 

Adding Class A and B NGC galaxies leads to an estimate of the fraction of significantly interacting galaxies among the bright part of the local galaxy populations of $71\pm8$ galaxies, or $4.8\%\pm0.5\%$. This is consistent with previous determinations, such as the number of interacting galaxies of around 4\% of bright local galaxies reported on the basis of a much smaller sample by Knapen \& James (2009). 

\subsection{Non-interacting very close companions}

  \begin{figure*}
   \centering
   \includegraphics[width=0.95\textwidth]{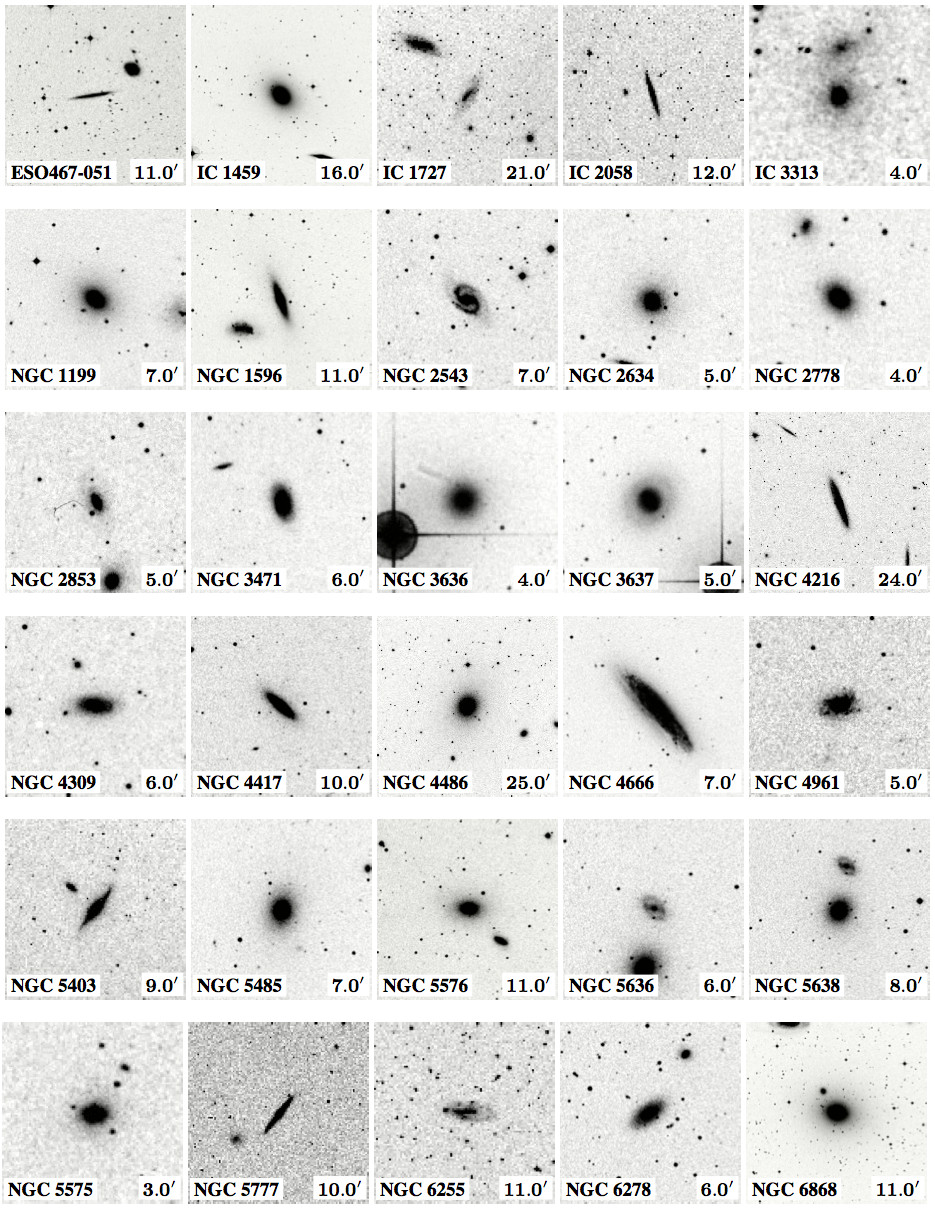}
   \caption{As Fig.\ref{ClassA}, now for galaxies with very close companions (closer than 1.5 times the diameter) but which do not show any indication of distorted morphology.}
              \label{Closenotinteracting}%
    \end{figure*}

    \setcounter{figure}{5}   
  \begin{figure*}
   \centering
   \includegraphics[width=0.95\textwidth]{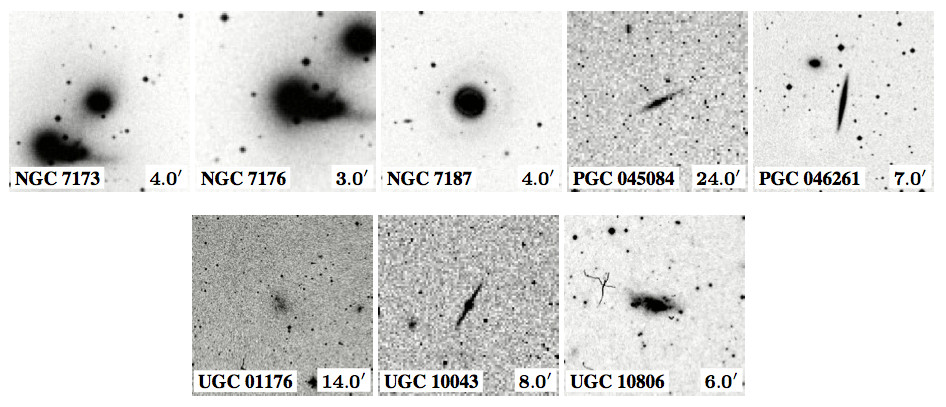}
   \caption{Continued.}
    \end{figure*}

\begin{table}[!h]
  \centering
  \caption{Galaxies with a very close companion but without any sign of interaction.}
  \begin{tabular}{lccc}
Galaxy & $D_{25}\,(')$  & Separation ($D_{25}$)  & Type\\
\hline\hline
  ESO 467-051   & 2.8     & 1.0     & SBdm sp  \\
  IC 1459       & 4.6     & 1.4     & E5(s)        \\
  IC 1727       & 6.5     & 1.2     & SB(s)m \\
  IC 2058       & 3.4     & 0.5     & SB(s)m sp \\
  IC 3313       & 1.3     & 0.8     & E0(s)            \\
  NGC 1199      & 2.8     & 1.2     & E3(s)           \\
  NGC 1596      & 3.9     & 0.8     & S0$^{-}$ sp \\
  NGC 2543      & 2.2     & 1.5     & SA\_B(s)b \\
  NGC 2634      & 1.7     & 1.1     & S0(s)            \\
  NGC 2778      & 1.3     & 1.4     & E3(s)            \\
  NGC 2853      & 1.7     & 1.3     & E5(s)            \\
  NGC 3471      & 1.7     & 1.4     & SB(s)0a \\
  NGC 3636      & 2.5     & 1.5     & S0(s)        \\
  NGC 3637      & 2.5     & 1.5     & SB\_a0$^+$  \\
  NGC 4216      & 7.8     & 1.5     & SAB\_a \\
  NGC 4309      & 1.9     & 0.8     & SAB0$^{\rm o}$ \\
  NGC 4417      & 3.1     & 0.2     & S0(s)        \\
  NGC 4486      & 7.1     & 1.2     & E0(s)        \\
  NGC 4666      & 5.0     & 1.5     & SB(s)c sp \\
  NGC 4961      & 1.1     & 1.3     & SB(rs) bc \\
  NGC 5403      & 2.8     & 0.6     & Sab \\
  NGC 5485      & 2.5     & 1.5     & E0(s)        \\
  NGC 5576      & 2.8     & 1.0     & SA0/a or E3 \\
  NGC 5636      & 1.4     & 1.4     & SAB(r)0/a \\
  NGC 5638      & 1.9     & 1.0     &         \\
  NGC 5775      & 3.7     & 1.2     & Sc sp \\
  NGC 5777      & 3.0     & 0.9     & SB0$^+$ \\
  NGC 6255      & 3.1     & 0.4     & SB(s) d \\
  NGC 6278      & 1.7     & 1.4     & SA(r)0$^{\rm o}$ \\
  NGC 6868      & 3.6     & 0.4     & E3(s)            \\
  NGC 7173      & 1.9     & 0.7     & E0(s)            \\
  NGC 7176      & 1.1     & 0.4     & S0(s)        \\
  NGC 7187      & 1.1     & 0.1     & S0a(r)        \\
  PGC 045084    & 2.3     & 1.4     & SB(s)a \\
  PGC 046261    & 2.5     & 0.7     & Scd sp \\
  UGC 01176     & 3.9     & 1.5     & Ia \\
  UGC 10043     & 2.2     & 1.2     & S0/a  \\
  UGC 10806     & 2.1     & 0.4     & SB(s)m \\
\hline\hline
  \end{tabular}
\tablefoot{Properties of galaxies with a companion closer than $1.5\,D_{25}$, but in which no signs of interaction are seen in our images. Morphological type from the RC3. Properties of the companion galaxies are in Table~3.}
\label{close-table}
\end{table}
     
We have identified an interesting class of galaxies, namely those with very close companions (closer than 1.5 times the diameter of the sample galaxy) but for which we cannot see any indication of distorted morphology. These 38 galaxies are shown in Fig.~\ref{Closenotinteracting} and listed in Table~\ref{close-table}. Most of these galaxies are of early morphological type, with only five spirals later than type Sb, and only one irregular galaxy (UGC~1176). We thus postulate that most of the galaxies in this category have rather stable, gas-poor disks, which are less susceptible to be distorted by gravitational interactions. They do constitute a curious class of galaxies that warrants further study.


\section{Conclusions}

   We have collected and re-reduced optical images of 1768 of the 2829 galaxies in the sample of the S$^4$G survey, which we make available to the community as ready-to-use FITS files to be used in conjunction with the mid-IR images. Of these, 1657 were observed as part of the SDSS. We collected and re-processed images in five bands for these galaxies, producing mosaics which cover an area of at least 5\,arcmin or three times the diameter of the galaxy, whichever is largest. 
   
   In addition, we observed in the $g$-band an additional 111 S$^4$G galaxies in the northern hemisphere with the 2.0\,m LT, so that optical imaging is released for 1768 galaxies, or for 63\% of the S$^4$G sample. We also produce false-colour images of some of these galaxies to be used for illustrative and public outreach purposes.

   We note from these images and from a NED database search which of the S$^4$G sample galaxies have close companions or are interacting, confirming this by determining the radial velocities, projected distances and magnitudes of the galaxies. We find that 17\% of the S$^4$G galaxies (21\% of those brighter than 13.5\,mag) have a close companion (within a radius of five times the diameter of the sample galaxy, a recession velocity within $\pm200$\,km\,s$^{-1}$ and not more than 3 mag fainter), and that around 5\% of the bright part of the S$^4$G  sample show significant morphological evidence of interaction. This confirms and further supports previous estimates of these fractions.
 
    The images described in this paper, the re-processed SDSS ones, the new LLT images, and the false-colour pictures, are publicly released for general use for scientific, illustrative, or public outreach purposes. The galaxies with close companions and, in particular, the interacting galaxies we identified are candidates for further study, either collectively or as case studies.

 
\begin{acknowledgements}

 We thank S\'ebastion Comer\'on, Benne Holwerda, and an anonymous referee for valuable comments on an earlier version of this paper, and Seppo Laine for discussions. We thank the entire S$^4$G team for conceiving, implementing and executing the S$^4$G survey. JHK, MC and RL acknowledge financial support to the DAGAL network from the People Programme (Marie Curie Actions) of the European Union's Seventh Framework Programme FP7/2007-2013/ under REA grant agreement number PITN-GA-2011-289313. JR acknowledges the receipt of a summer studentship at the Instituto de Astrof\'\i sica de Canarias. JHK, SE-F, JR, JB, MC and RL thank co-author NS for his dedication, time, and effort in collaborating with us in turning FITS images into inspiring and colourful pictures of galaxies---this constitutes another successful Pro-Am collaboration. The Liverpool Telescope is operated on the island of La Palma by Liverpool John Moores University in the Spanish Observatorio del Roque de los Muchachos of the Instituto de Astrof\'\i sica de Canarias with financial support from the UK Science and Technology Facilities Council. Funding for SDSS-III has been provided by the Alfred P. Sloan Foundation, the Participating Institutions, the National Science Foundation and the US Department of Energy Office of Science. The SDSS-III web site is http://www.sdss3.org/. SDSS-III is managed by the Astrophysical Research Consortium for the Participating Institutions of the SDSS-III Collaboration including the University of Arizona, the Brazilian Participation Group, Brookhaven National Laboratory, University of Cambridge, Carnegie Mellon University, University of Florida, the French Participation Group, the German Participation Group, Harvard University, the Instituto de Astrof\'\i sica de Canarias, the Michigan State/Notre Dame/JINA Participation Group, Johns Hopkins University, Lawrence Berkeley National Laboratory, Max Planck Institute for Astrophysics, Max Planck Institute for Extraterrestrial Physics, New Mexico State University, New York University, Ohio State University, Pennsylvania State University, University of Portsmouth, Princeton University, the Spanish Participation Group, University of Tokyo, University of Utah, Vanderbilt University, University of Virginia, University of Washington and Yale University. This research has made use of NASA's Astrophysics Data System. We acknowledge the usage of the HyperLeda data base and the NASA/IPAC Extragalactic Data base (NED), operated by the Jet Propulsion Laboratory, California Institute of Technology, under contract with the National Aeronautics and Space Administration. We have used data products made available through the Galaxy Zoo collaboration (www.galaxyzoo.org). The Digitized Sky Survey was produced at the Space Telescope Science Institute under US Government grant NAG W-2166. 

\end{acknowledgements}




\begin{longtab}
\label{companionstable}
\begin{longtable}{lcccccccc}
\caption{Properties of galaxies with close companions, and of interacting galaxies.}\\
\hline\hline
Galaxy		& Inter.	& $D_{25}$ ($'$)	& Vel. (km/s)	 & mag. 	& \multicolumn{3}{c}{Separation} & Optical	 \\
			&		& 				&						 & 			& ($'$)	& $D_{25}$& (kpc) 	   & \\
\hline
\endfirsthead
\caption{Continued.}\\
\hline\hline
Galaxy		& Inter.	& $D_{25}$ ($'$)	& Vel. (km/s)	 & mag. 	& \multicolumn{3}{c}{Separation} & Optical	 \\
			&		& 				&						 & 			& ($'$)	& $D_{25}$& (kpc) 	   & \\
\hline
\endhead
\hline
\endfoot
\textbf{ESO 027-001               }	&  	&  2.7	& 2570 & 	12.2	&	&	&	&  \\
\quad   ESO 027- G 003               	& 	&  	& 2578 & 	14.8	&	   7.9 &	2.9&	78.4 &  \\
\textbf{ESO 079-003}	&  	&  3.0	& 2616 & 	12.6	&	&	&	&  \\
\quad   ESO 079- G 002               	& 	&  	& 2756 & 	13.7	&	   8.2 &	2.7&	87.4 &  \\
\textbf{ESO 185-027}	&	&  1.4& 2167 & 13.8	&	&	&	& DR8 \\
   \quad   6dF J1956175-565025                	& 	&  	& 2112 & 	16.3	&	   4.2 &	3.1&	34.5 &  \\
\textbf{ESO 236-039}	& B	&  1.0	& 1574 & 	15.1	&	&	&	&  \\
\quad   AM 2141-491 NED01 SE         	& 	&  	& 1546 & 	15.6	&	   3.1 &	3.1&	18.4 &  \\
\hline

\end{longtable}
 
\tablefoot{Galaxies with close companions. Column 1: name. Galaxies in boldface are S$^4$G sample galaxies. Column 2 indicates whether galaxies show signs of interaction (see Sect.~6.2). Column 3 is the diameter of the galaxy, in arcmin. Column 4 and 5 are the recession velocity and optical magnitude, respectively.  Columns 6-8 are the projected separation between the companion and the sample galaxy, in units of arcmin, sample galaxy diameter, and kpc, respectively. The final column, 9, indicates whether we release optical imaging for a galaxy, and from which source. All name, position, diameter, velocity and magnitude data as retrieved from the NED. A machine-readable version of the table is available from the CDS.}

\end{longtab}



\end{document}